\documentstyle[12pt]{article}
\begin{document}
\author{Dr.~Kevin Leung$^{[a]*}$, Dr.~Alexander J.~Pearse,$^{[b]}$ 
Dr.~A.~Alec~Talin,$^{[c]}$ \\
Dr.~Elliot J.~Fuller,$^{[c]}$ Prof.~Gary
W.~Rubloff,$^{[b]}$ and Dr. Normand~A.~Modine$^{[a]}$}
%\date{\today}
\title{Kinetics-Controlled Degradation Reactions at Crystalline
LiPON/Li$_x$CoO$_2$ and Crystalline LiPON/Li-metal Interfaces}
\maketitle

\centerline{$^{[a]}$Sandia National Laboratories, Albuquerque, NM 87185}
\centerline{email: {\tt $^*$kleung@sandia.gov}; fax: 15058445470} 
\centerline{$^{[b]}$Department of Materials Science and Engineering,
	University of Maryland} 
\centerline{$^{[c]}$Sandia National Laboratories, Livermore, CA 94550} 
\input epsf
%\ssp

\renewcommand{\thetable}{\arabic{table}}

\begin{abstract}

Detailed understanding of solid-solid interface structure-function relations
is critical for the improvement and wide deployment of all solid-state
batteries.  The interfaces between lithium phosphorous oxynitride (``LiPON'')
solid electrolyte material and lithium metal anode, between LiPON and
Li$_x$CoO$_2$ cathode surfaces, have been reported to generate solid
electrolyte interphase (``SEI'')-like products and/or disordered regions.
Using electronic structure calculations and crystalline LiPON models with
atomic-layer-deposition-like stoichiometry, we predict LiPON models with
purely P-N-P backbones are kinetically inert towards lithium at room
temperature.  In contrast, transfer of oxygen atoms from low-energy
Li$_x$CoO$_2$ (104) surfaces to LiPON is much faster under ambient
conditions.  The mechanisms of the primary reaction steps, LiPON motifs
that readily react with lithium metal, experimental results on amorphous LiPON
to partially corroborate these predictions, and possible mitigation strategies
to reduce degradations are discussed.  LiPON interfaces are found to be useful
case studies for highlighting the importance of kinetics-controlled processes
during battery assembly at moderate processing temperatures.

\vspace*{0.5in}
\noindent keywords: lithium ion batteries; LiPON;
{\it ab initio} calculations; solid-solid reactions; interfaces.

\newpage

\end{abstract}

\section{Introduction} \label{intro}

The deployment of solid electrolytes in all-solid-state batteries in
transportation energy storage applications can effectively address safety
concerns associated with current commercial, organic-solvent-based lithium
ion batteries.  Technical issues
remain\cite{review0,review1,review2,review3,review4,review5}, many of which
are associated with buried solid-solid interfaces between electrolytes and
electrodes\cite{interface}.  While cross-sectional transmission electron
microscopy (TEM)\cite{meng,meng1,meng2},
potential mapping\cite{holography1,holography2}, X-ray photoemission
spectroscopy\cite{jaegermann10,jaegermann11,jaegermann15,jaegermann_rev,hausbrand},
and other experimental techniques have provided a wealth of information about
such interfaces, so far they lack sufficient resolution to yield atomic
length-scale details -- especially in materials without crystalline order.

Electronic structure (e.g., Density Functional Theory, DFT) calculations
on model solid-solid interfaces yield insights complementary to 
measurements\cite{tateyama1,tateyama2,holzwarth,holzwarth1,dallas,sumita,sumita1,murugan,liponsurdft,albe2017}.
They can further raise novel scientific questions that will attract experimental
inquiry.  One important topic to address is the charge separation associated
with the solid-state electric double layer (EDL) at such
interfaces\cite{jaegermann11b}.
Electric field associated with the EDL are expected to aid Li$^+$ diffusion
during charge and discharge events, but may also accelerate interfacial
chemical or electrochemical reactions between electrode and electrolyte to
form a ``solid electrolyte interphase'' (SEI)\cite{solid}.  Such
SEI products have indeed been reported at some all-solid battery interfaces.
Liquid electrolytes also take part in interfacial reactions, although
in that case the SEI is often formed during the first few charging cycles,
while in all-solid-state batteries part of the SEI already emerges during
battery assembly\cite{note1}.

This work focuses on crystalline model of lithium phosphorous oxynitride
(``LiPON'')\cite{meng,jaegermann10,jaegermann15,hausbrand,rubloff,talin,talin1,lipon3,lipon4,lipon5,lipon6,lipon_coat_lco,exposed,amy,n3effect,lipon_li,lipon_lco,janek} solid
electrolyte materials in contact with lithium metal (Li(s)) anode and lithium
cobalt oxide (Li$_x$CoO$_2$) cathode surfaces.  LiPON interfaces present
interesting case studies.  It has been reported that LiPON forms interfacial
reaction products when in contact with Li(s)\cite{jaegermann15} and with
Li$_x$CoO$_2$\cite{meng,meng1,meng2,jaegermann10,hausbrand}, especially
at elevated temperature, although the batteries continue to function.  Heating
may even reduce interfacial charge transfer resistance\cite{ogumi2005}.
Therefore modeling of the electrode/electrolyte interfaces, which control
the EDL and charge transfer, necessarily requires first addressing the
reactions that produce SEI.  Here we distinguish SEI products obtained from
gaseous precursors during battery fabrication and assembly, before the solid
component being grown is fully formed, and SEI products which emerge during
cycling\cite{note1}. The latter register as increase in SEI after cycling,
and are the focus of our modeling work.

We apply DFT calculations to study LiPON interfacial reactions on the low energy
Li(001) and Li$_x$CoO$_2$ (104) surfaces.  While many solid-state batteries
feature silicon anodes\cite{lipon3}, the top layer of lithiated silicon is
typically terminated by Li atoms\cite{fec}, which minimizes the surface energy.
Therefore our work may also be relevant to Si anodes.  On the cathode side,
our calculations draw on previous modeling work on Li$_x$CoO$_2$ bulk
crystal\cite{cedersur}, its (vacuum) surfaces\cite{cedersur,mengsur,giordano},
electron spin distributions,\cite{meng,sumita} and on LiPON studies using model
structures\cite{liponsurdft,holzwarth_bulk,holzwarth_bulk1,n3jcp,lipon_a}.
Related modeling work on Li(Mn,Ni,Co)O$_2$ surfaces are also
relevant\cite{morgan2017,iddir2017}.
Our work also benefits from single phase thermodynamics predictions\cite{mo},
which describe the stability window outside which LiPON
and electrode materials can react.  If reactions are not limited by slow
kinetics and reach equilibrium, single phase thermodynamics efficiently
predict the final, most stable interfacial products.  However, unlike crystal
growth associated with cathode synthesis routinely conducted at
700-1000$^o$C, the processing temperatures for fabricating interfaces are much
lower (150-300$^o$C for oxides, and lower for sulfides).  So metastable
products may dominate at solid-solid
interfaces.  Indeed, there is experimental evidence that kinetics- rather
than thermodynamics-determined products are formed in some all-solid
batteries\cite{meng,gewirth}.  Even more obvious examples are the anode SEI
in organic liquid electrolyte-based batteries.  The SEI films there are formed
at room temperature; many SEI components are demonstrably metastable
and exist only because of complex kinetics constraints\cite{batt}. In 
liquid-electrolyte SEI, single-phase thermodynamics predictions clearly fail;
combining kinetics and thermodynamics modeling is crucial to yield insights
that can be related to measurements.

Here we apply perspectives from liquid state batteries\cite{lmo,chemmat,batt}
and examine the activation energies of rate-limiting primary degradation
reactions at explicit electrode-electrolyte interfaces.  A few subsequent,
secondary reactions are also considered; in some cases, they are faster than
the first reactions which activate the chain of degradation events.  Unlike
single-phase thermodynamic calculations\cite{mo}, this kinetic approach does not
predict the final products, which may be the culmination of reactions many steps
later.  Instead, we examine the temperature and voltage dependence of primary
reaction steps, elucidate the mechanisms, and gain insight into what material
variations can raise reaction barriers.  This will allow mitigation of
degradation processes.  We predict faster degradation reactions on the cathode
than on the anode surface, but some reactions can occur at room temperature
at both interfaces.

For experimental work, we perform conductivity measurements on pristine and
LiPON-coated Li$_x$CoO$_2$ to support the prediction that transfer of chemical
species between the LiPON surface layers and Li$_x$CoO$_2$ occurs.  Finally,
X-ray photoelectron spectroscopy (XPS) is used to confirm chemical changes at
the cathode/LiPON interface.

There are many challenges associated with the computational approach stated
above.  LiPON encompasses a family of amorphous materials synthesized with
different deposition techniques.  Different LiPON materials exhibit variations
in chain lengths and stoichiometries, especially in the nitrogen content.  We
start with a crystalline Li$_2$PO$_2$N model\cite{holzwarth_bulk} which is
close to the stoichiometry obtained in atomic layer deposition (ALD)
synthesis\cite{lipon3}.  The model has infinitely long chains, exclusively
2-coordinated N atoms, and no P-O-P or P(N)$_3$ motifs, although the latter
are minority LiPON features under low temperature growth
conditions\cite{jaegermann10,jaegermann11,lipon4}.  We also consider defected
LiPON modified from perfect LiPON crystals.  We stress that our modeling
work is not meant to match any particular set of measurement.  Instead,
we elucidate interfacial reaction mechanisms and kinetically stable/reactive
structural motifs that should be of general interest, and inform
interpretations about the reactions of different LiPON realizations.

Atomic structures normally used as starting point of modeling efforts are
largely unknown from existing measurements of buried battery interfaces.
Registry- and lattice-matching between two solid state components are
non-trivial; the need to account for possible addition or removal of Li
atoms at battery interfaces adds complexity.  As interfaces usually involve
spatial heterogeneities, the predicted reaction rates may span a continuum.
Therefore multiple reactive sites are examined.   The Co charge- and
spin-states at the LiPON interface can change with the coordination
environment.  The reproducibility of the predicted set of Li$_x$CoO$_2$
spin states is an issue.  The literature mostly focus on either
$x$=0 or $x$=1 surfaces\cite{cedersur,mengsur,giordano}, which exhibit
all Co$^{3+}$ or Co$^{4+}$, circumventing this issue.  We show that
the problem must be confronted for slab geometries at intermediate $x$ values.
Regarding voltage-dependence\cite{solid,arias,neurock,rossmeisl15,qi_acc_chem},
reactions at interfaces may or may not vary with applied potential, depending
on whether the rate-limiting step involves electron transfer or not; this
distinction will be examined.  Qualitative comparison between our predictions
and experimental results in the literature is presented in the Discussion
Section.  

\section{Results: LiPON-Lithium Interfaces}

\subsection{LiPON slab on Li(s) surface}

\begin{figure}
\centerline{\hbox{ (a) \epsfxsize=2.00in \epsfbox{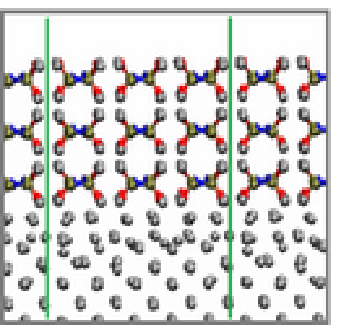} 
	           \epsfxsize=2.00in \epsfbox{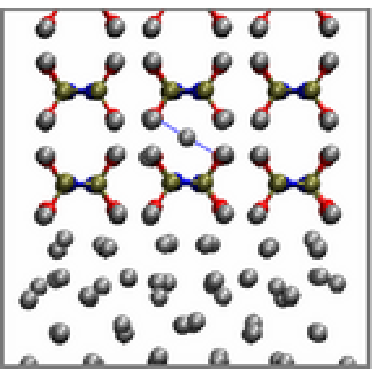} (b) }}
\centerline{\hbox{ (c) \epsfxsize=2.00in \epsfbox{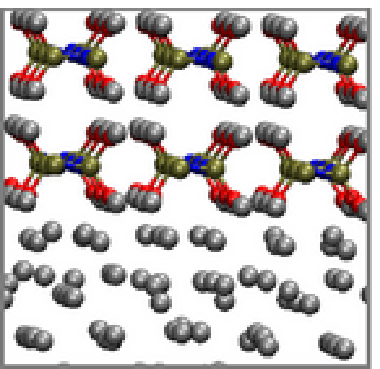}  
	           \epsfxsize=2.00in \epsfbox{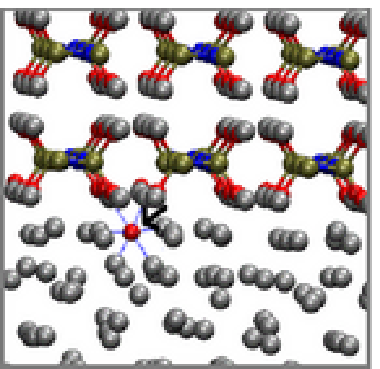} (d) }}
\caption[]
{\label{fig1} \noindent
(a): Periodically replicated simulation cell (green boundaries) containing
LiPON slab on Li metal.  A vacuum region exists. (b): Expanded view with an
extra Li atom in LiPON interior region.  (c)-(d): Interface region in (a)
expanded, before/after P-O bond-breaking, respectively.  Silver, red, blue,
and dark green represents Li, O, N, and P atoms, respectively.  Blue dashed
lines depict selected Li-O distances that are less than 2.2~\AA.  Black arrows
in some panels indicate the motion of key atoms in bond-breaking reactions.
}
\end{figure}

Reaction between LiPON and lithium metal to form Li$_3$P, Li$_2$O, and
LiPN$_2$ are thermodynamically favorable\cite{mo}, indicating
that LiPON is metastable against lithium.  It is the reaction rate 
that needs to be determined.

Figure~\ref{fig1}a depicts a simulation cell with model LiPON in a slab
geometry.  There are 18~LiPON O-atoms in contact with the Li metal.  As
discussed in the supporting information document (S.I.), one Li~atom is
inserted beneath each O$^{2-}$ to form
an interlayer.  Figures~\ref{fig1}c-d are expanded view of the reaction region,
before and after breaking one P-O bond.   In Fig.~\ref{fig1}d, after optimizing
the atomic configuration, the detached O$^{2-}$ anion becomes buried inside the
Li metal, coordinated to 6~Li with O-Li bond lengths of 2.1~\AA\, or less.
Bader charge analysis\cite{bader} is qualitatively consistent with the transfer
of two electron from Li metal to the original P$^{5+}$ to form
P$^{3+}$ with a broken bond.  The energy released is significant,
with $\Delta E$=-0.63~eV.

At the transition state, the P-O distance is 2.27~\AA.  The reaction barrier
$\Delta E^*$ is predicted to be a very substantial 2.15~eV.  Using an Arhenius
rate expression $1/\tau(T) $=$ k \exp(-\Delta G^*/k_{\rm B}T) $, approximating
the free energy barrier with the zero temperature $\Delta E^*$, and assuming
a kinetic prefactor of $k$=10$^{12}$/s, the room temperature reaction time
$\tau(T)$ associated with breaking a bond with $\Delta E^*$=0.92~eV is about
an hour; breaking a bond with $\Delta E^*>1.1$~eV takes far beyond battery
operational time scales.  At T$\approx$690~K, $\tau(T)$ shrinks to one hour.
However, this is above the lithium metal melting point.  Lithium interfaces
must be assembled below 500~K.  At this temperature, the predicted
$\Delta E^*$ is too still high for the reaction to occur within reasonable
timescales.  The high activation energy associated with P-O bond
cleavage is likely the reason Li$_3$PO$_4$ has also been used as coating
layers for lithium metal\cite{li3po4}, even though Li$_3$PO$_4$ is
also thermodynamically unstable against Li metal\cite{albe2017}.

We have also attempted to break the other 17 P-O bonds at the interface.
The $\Delta E$ predicted depends on the coordination of the released O$^{2-}$,
and ranges from -0.77~eV to -0.30~eV.  They average to -0.53$\pm$0.03~eV.
$\Delta E^*$ calculations are more costly, and are only attempted for two
other P-O cleavage events.  $\Delta E^*$=2.16 and 2.30~eV are found to be
associated with breaking these other bonds, again indicating that the
reaction would be very slow under room conditions.

Attempts to break single P-N bonds in a LiPON chain at the interface and then
re-optimize
the configuration lead to either reformation of the P-N bond and the original
LiPON structure, or a metastable structure with the P and N atoms separated
by 2.22~\AA\, (not shown) instead of the 1.55-1.59~\AA\, P-N bond length
in equilibrium LiPON models.  The energy of the configuration with a broken
bond is a very significant 1.76~eV above that of the intact LiPON slab.  While
breaking two P-N bonds and depositing the N atom onto the Li metal far from
the two P atoms is exothermic, we have not found a pathway with a sufficiently
low reaction barrier to justify this mechanism at room temperature.  Thus we
conclude that neither P-O nor P-N cleavage is kinetically viable at the
crystalline LiPON/Li metal interface.

\begin{figure}
\centerline{\hbox{ \epsfxsize=3.50in \epsfbox{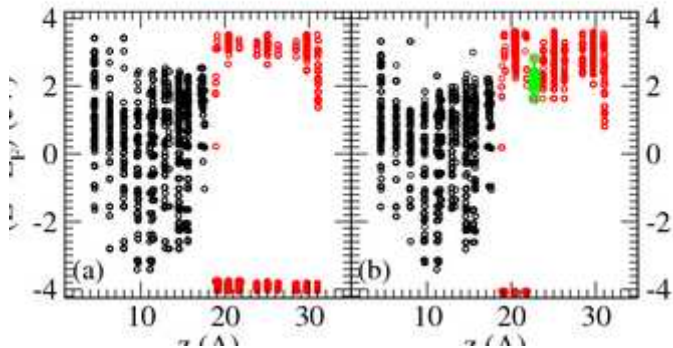} }}
\caption[]
{\label{fig2} \noindent
(a)-(b) Kohn-Sham orbitals corresponding to Fig.~\ref{fig1}a-b.
Black and red spheres correspond to location of orbitals along $z$-axis
(perpendicular to interface) localized on Li metal and LiPON atoms,
respectively.  Green circles correspond to orbitals on the excess Li
added in the LiPON interior region.  They lie above the Fermi level and
are unoccupied.
}
\end{figure}

Because the model system is metallic and the simulation cell has a vacuum
region, the instantaneous electrochemical potential or voltage ${\cal V}_e$ of
this system can be unambiguously assigned.  ${\cal V}_e$ is the work function
$\Phi$ divided by $|e|$ and subtracted by 1.37~V\cite{solid}.  Before 
breaking the P-O bond, ${\cal V}_e$=0.12~V vs.~Li$^+$/Li(s) reference.  
The orbital alignment is depicted in Fig.~\ref{fig2}a.  Unlike the
calculations in Refs.~\cite{lipon_a,albe2017}, Fig.~\ref{fig2}a accounts for
explicit LiPON/lithium anode interfaces; although this figure represents thin
films, absolute (not relative) orbital energy levels can be obtained.

${\cal V}_e$ is sufficiently close to the equilibrium voltage below which Li
metal is stable, at 0.0~V vs.~Li$^+$/Li(s), that the predicted $\Delta E$
and $\Delta E^*$ can be regarded as associated with approximately the
equilibrium Li-plating or stripping voltage.  If P-O bond breaking in
Fig.~\ref{fig1}d is a electrochemical reaction, according to the Butler-Volmer
equation\cite{butler}, $\Delta E^*$ is expected to be lowered by 0.12~eV
relative to the same reaction at 0.0~V, assuming that the rate-determining 
step involves two-electron transfer and $\alpha=0.5$\cite{butler}.
This reduction in $\Delta E^*$ would not change the conclusion that 
$\Delta E^*$ remains far too high for P-O or P-N breaking reaction to occur
at room temperature.  We will show below that, in fact, barriers associated
with LiPON reactions on lithium metal do not strongly depend on ${\cal V}_e$.

It is also of interest to observe whether long-range electron injection 
into defects in the middle of the LiPON solid region can occur.  In batteries
based on liquid electrolytes, such $e^-$ transfer has been widely acknowledged.
To this end, we have attempted to break a P-N bond in the LiPON crystal region
away from the interface, with one or two Li inserted in the void space around
the broken bond, while maintaining charge neutrality.  Such a ``grand
canonical'' scheme represents reactions accompanied excess Li migration into
LiPON from Li metal.  After optimization of the atomic configuration, the P-N
bond is reconstituted (Fig.~\ref{fig1}b), and Bader charge analysis indicates
negligible excess $e^-$ residing in the Li inserted. This finding is consistent
with the predicted orbital alignment (Fig.~\ref{fig2}b).  In other words, the
inserted Li is a Li$^+$.  The total energies of these systems are less
favorable than without Li insertion into the LiPON interior region by 1.16~eV
after accounting for the Li metal cohesive energy.  Note that ${\cal V}_e$
assicated with Fig.~\ref{fig1}d and Fig.~\ref{fig2}b is 0.28~V.
The slight difference in ${\cal V}_e$ compared with that in Fig.~\ref{fig1}c
is due to the change in the surface dipole moment following the insertion.
It is an artifact of our finite simulation cell.  Although ${\cal V}_e$ is
found to be slightly too high to correspond to the true 0.0~V vs.~Li$^+$/Li(s),
we can at least conclude that at ${\cal V}_e$ $\sim$0.12~to~0.28~V, long range
electron transfer into the interior of our LiPON model should not occur.

\subsection{LiPON chain or fragment on Li(s) surface}

\begin{figure}
\centerline{\hbox{ (a) \epsfxsize=2.00in \epsfbox{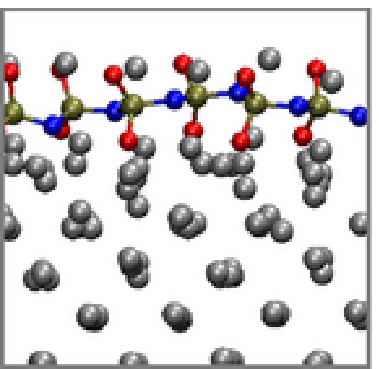}  
	           \epsfxsize=2.00in \epsfbox{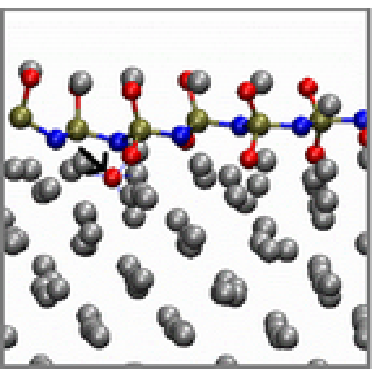} (b) }}
\centerline{\hbox{ (c) \epsfxsize=2.00in \epsfbox{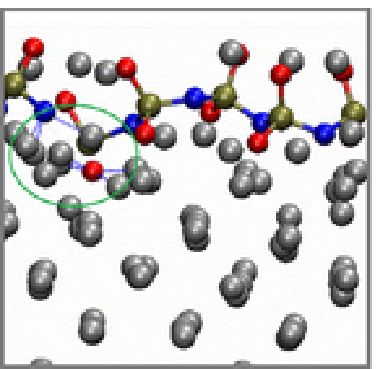}  
	           \epsfxsize=2.00in \epsfbox{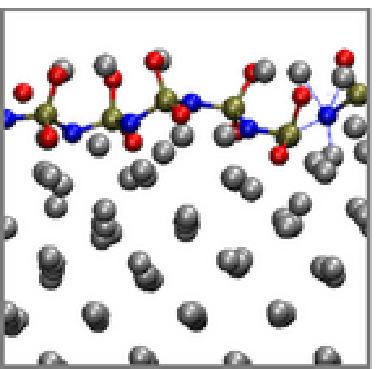} (d) }}
\centerline{\hbox{ (e) \epsfxsize=2.00in \epsfbox{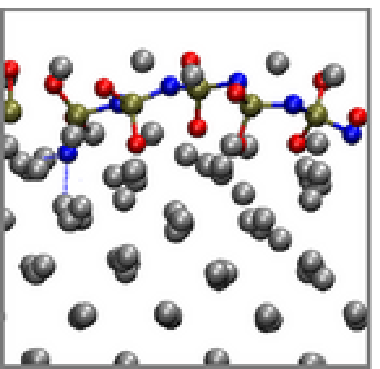}  
	           \epsfxsize=2.00in \epsfbox{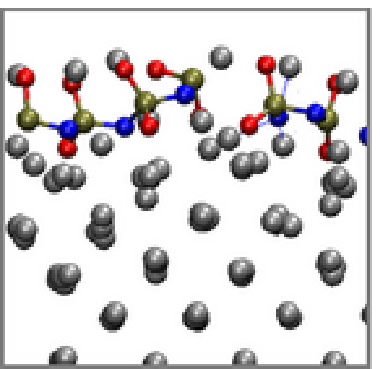} (f) }}
\caption[]
{\label{fig3} \noindent
Simulation cell with LiPON chain on Li metal.  (a): intact; (b): broken
P-O bond; (c): attempting to calculate barrier to (b) breaks a P-N bond
instead; (d)-(f): broken P-N bond with edge N-atom coordinated to
4, 2, and 3 Li, respectively.  For color scheme, see Fig.~\ref{fig1}.
}
\end{figure}

The previous section fails to explain why LiPON decomposes on Li metal
surfaces at room temperature\cite{jaegermann15}.  Next we consider the fact
that LiPON is not crystalline.  It consists of short chains in disordered
environments\cite{lipon_a}.  We attempt to model such regions as a single
Li$_2$PO$_2$N chain deposited on Li metal.  Fig.~\ref{fig3}a is obtained by
removing all other periodically replicated chains in Fig.~\ref{fig1}a, and
re-optimizing the atomic configuration.  This lifts structural constraints
(e.g., hindrance of local rotation along the P-N-P chain) which stabilize
LiPON against bond breaking.

There are 12 P-N bonds in the LiPON backbone in the simulation cell,
6 P-O bonds with O atoms in contact with the metal surface, and
another 6 P-O bonds with O atoms pointing outwards (Fig.~\ref{fig3}a).
Fig.~\ref{fig3}b depicts a configuration where one of the first group of
P-O bonds is broken.  The reaction results in $\Delta E$=-1.15~eV.  The
products are more exothermic than those obtained by breaking most 
P-O bonds in the crystalline LiPON slab (Fig.~\ref{fig1}b).  Our attempt to
compute a reaction barrier for this reaction leads to breaking a P-N bond
in addition to the initial P-O cleavage event (Fig.~\ref{fig3}c), yielding
a very exothermic $\Delta E$=-2.41~eV relative to the initial intact LiPON
chain.  This suggests that breaking P-N bonds in a LiPON chain exhibits
lower barriers than breaking P-O, and that P-O cleavage intermediates are
unlikely to be the first reaction products.

Next we focus on the twelve P-N bonds.  Fig.~\ref{fig3}d depicts the optimzed
configuration after breaking one of these bonds.  $\Delta E$ is $-2.04$~eV.
The reaction is highly exothermic, in contrast to the analogous reaction in
the slab geometry discussed in the last section.  The terminal N-atom generated
by the broken bond ``burrows'' into the Li metal region, becoming coordinated
to 4~Li atoms (Fig.~\ref{fig3}d).  The LiPON chain undergoes significant
conformational changes to accommodate this motion.  Such motions
are hindered in a LiPON crystal environment (Fig.~\ref{fig1}a).

However, $\Delta E^*$=1.63~eV is predicted.  It is smaller than P-O bond
cleavage $\Delta E^*$ described in the last section, but remains too large for
a room temperature, one-hour reaction time scale.  At the transition state, the
P-N distance is 2.21~\AA.  Incidentally, the combination of large negative
$\Delta E$ and large positive $\Delta E^*$ is not exceptional.  This is one
of many examples where a large ``thermodynamic driving force'' is correlated
with a slow reaction rate.  It shows that $\Delta E$ and $\Delta E^*$ are
not necessarily correlated, and that reaction barriers must be explicitly
computed to understand interfacial reactions.  

We have also attempted to break the other eleven P-N bonds, one by one.  Two
attempts revert back to the intact LiPON chain structure.  The other nine
yield exothermic reactions.  The $\Delta E$ of the resulting metastable
structures vary from -0.87~eV to -1.32~eV, depending on whether the
N-atom broken off from the phosphorus is 4- (Fig.~\ref{fig3}d), 2-
(Fig.~\ref{fig3}e), or 3-coordinated (Fig.~\ref{fig3}f).  They average to
$\Delta E$=-1.20~eV$\pm$0.10~eV.  Given the low barriers associated with Li
motion on its metal surfaces, subsequent atomic configuration rearrangement
from the less exothermic ($\sim$-0.87~eV) to very exothermic ($\sim$-1.63~eV)
metastable product configurations via Li diffusion on the metal surface may
occur readily, but this is not the object of our studies.  Instead, we focus
on the barrier associated with primary P-N bond cleavage reaction, which
should be the rate-limiting step.  We have computed the reaction barriers of
four of the ten P-N cleavage reactions, in addition to the $\Delta E^*$ 
associated with $\Delta E$=-1.63~eV.  They average to $\Delta
E^*$=1.54$\pm$0.15~eV.  The barriers are too high to permit room
temperature reactions at reasonable timescales.

Experimentally, LiPON is known to be composed of finite chains.  In Sec.~S3
of the S.I., we consider short LiPON fragments instead of infinite chains.  
Those calculations suggest that the terminal P-O and P-N groups of LiPON
fragments, bonded to four-coordinated P$^{5+}$ atoms, are about as
kinetically stable as interior bonds.  

\subsection{Effects of Anode Voltage on LiPON chain}

\begin{figure}
\centerline{\hbox{ (a) \epsfxsize=2.00in \epsfbox{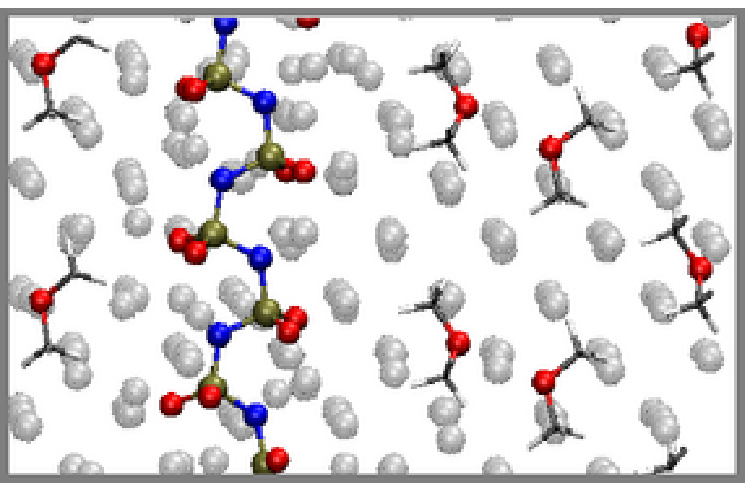} 
	           \epsfxsize=2.00in \epsfbox{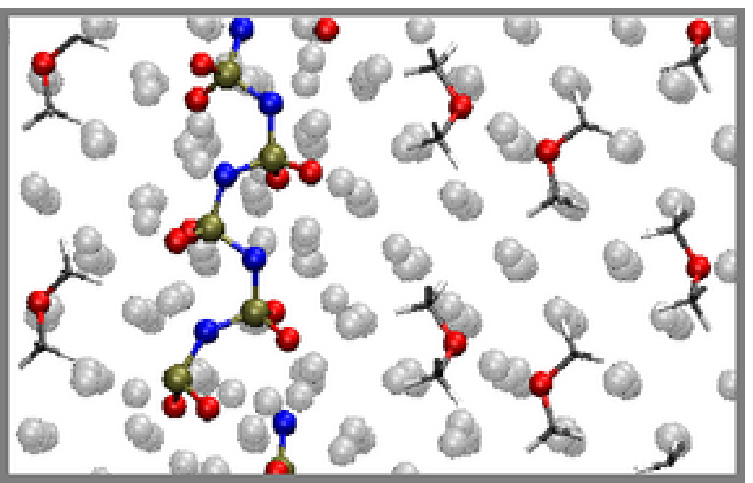} (b) }}
\caption[]
{\label{fig4} \noindent
(a)-(b) Intact and decomposed LiPON chain on enlarged Li(100) surface
simulation cell with 4 [(CH$_3$)$_2$O]$_2$Li$^+$ units which lower the
voltage.  The LiPON configurations are similar to Fig.~\ref{fig3}c-d.
For color scheme, see Fig.~\ref{fig1}.
}
\end{figure}

The instantaneous electronic ${\cal V}_e$ for the reactant configuration
in Fig.~\ref{fig3}a is predicted to be 0.57~V.  At this potential, Li metal
should dissolve into the electrolyte as Li$^+$, releasing $e^-$.  In other
words, the calculations associated with Fig.~\ref{fig3} are not at
electrochemical equilibrium\cite{solid}, but reflect a computational
overpotential of $\Phi$=0.57~V.  To lower the voltage to the equilibrium value
of 0~V, we expand the $x$-lattice constant from 16.41~\AA\, to 23.06~\AA.  The
simulation cell is created by taking the adsorbed LiPON geometry in
Fig.~\ref{fig3}a and adding a strip of bare Li (001) slab.  Therefore the
LiPON adsorption geometry should be unchanged.  Next, we
add 4~[(CH$_3$)$_2$O]$_2$Li$^+$ units adsorbed on the Li (001) surface away
from the LiPON chain (Fig.~\ref{fig4}a-b).  Ether ((CH$_3$)$_2$O) molecules 
are chosen because they are kinetically stable against lithium metal and do
not decompose during optimization calculations.  In the charge-neutral
simulation cell, the positive charges of the 4~added, ether-coordinated Li$^+$
induce negative surface charges on the Li metal surface.  This results in a
surface dipole density that reduces ${\cal V}_e$ to 0.04~V\cite{solid}.  On
this surface, breaking one of the P-N bonds (Fig.~\ref{fig4}b)
now yields $\Delta E$=-1.59~eV and $\Delta E^*$=1.61~eV.

Fig.~\ref{fig3}d, associated with breaking the same bond at
${\cal V}_e$=0.57~V in a smaller simulation cell, have -2.04~eV and 1.63~eV
for these values.  Despite the decrease of ${\cal V}_e$, expected to increase
the exothermicity of an electrochemical reaction, $\Delta E$ does not become
more negative.  The smaller $|\Delta E|$ magnitude for Fig.~\ref{fig4}b
compared with Fig.~\ref{fig3}b may arise from subtle reorientation of the
[(CH$_3$)$_2$O]$_2$Li$^+$ units.  $\Delta E^*$ is almost unchanged.  This
voltage-independence suggests that the rate determining step of the reaction
``does not involve'' $e^-$ transfer from the Li electrode.  This conclusion is
surprising.  Many charge transfer reactions, like water reduction or
oxidation on metal surfaces, exhibit $\Delta E$ that vary linearly
with overpotential $\Phi$.  In DFT calculations, the voltage dependence
is often added {\it a posteriori} to DFT values as $(ne \Phi)$, where $n$ is
the number of electrons transferred per reaction,\cite{norskov} and $e$
is the electronic charge.   Regarding
$\Delta E^*$, the Butler-Volmer equation assumes modification of the reaction
barrier by $n \alpha \Phi$, where $\alpha$ is typically 0.5\cite{butler}.

The fact that neither $\Delta E$ nor $\Delta E^*$ becomes more negative with
decreasing $\Phi$ suggests that the bond-breaking in LiPON on lithium metal is
a surface chemical, not electrochemical, reaction.  In metal surface catalyzed
electrochemical reactions\cite{norskov}, the electrons transferred to the
solution induce a delocalized surface charges on the metal electrode.  In the
reaction between Li(100) and the LiPON chain, we propose that two Li atom
on the metal surface turn into discrete Li$^+$, with their $e^-$ donated to
the LiPON chain and no delocalized positive charge induced on the electrode
surface.  In this way, the value of ${\cal V}_e$ associated with
Fig.~\ref{fig3}~and~\ref{fig4} do not affect the reaction energetics.  Note
that it is difficult to quantify the charge distribution on lithium metal 
surfaces.  Bader charge decomposition yields ambiguous, non-uniform charge
distributions even on pristine Li (100).

\subsection{Variations in LiPON Backbone}

The above calculations have not yet explained the small amount of SEI formation
observed upon depositing Li on pre-formed LiPON\cite{jaegermann15}.  Next we
turn to chemical variations along the LiPON chain.  LiPON is an amorphous solid
with variable chemical compositions\cite{review0}.  Many experimental papers,
as well as some modeling work\cite{n3jcp,lipon_a}, have shown that some
P-O-P linkages exist.

\begin{figure}
\centerline{\hbox{ (a) \epsfxsize=2.00in \epsfbox{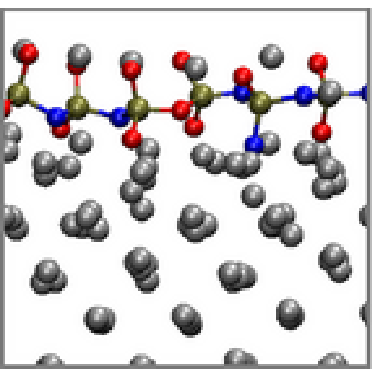} 
	           \epsfxsize=2.00in \epsfbox{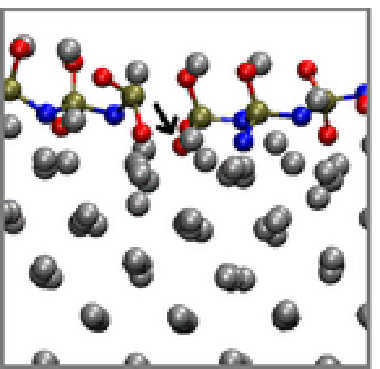} (b) }}
\centerline{\hbox{ (c) \epsfxsize=2.00in \epsfbox{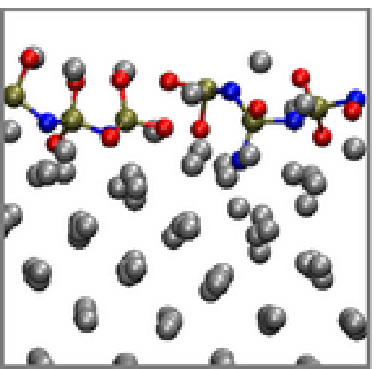} 
	           \epsfxsize=2.00in \epsfbox{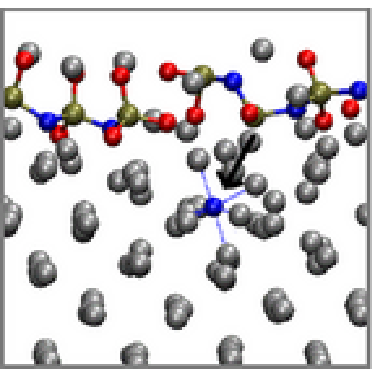} (d) }}
\centerline{\hbox{ (e) \epsfxsize=2.00in \epsfbox{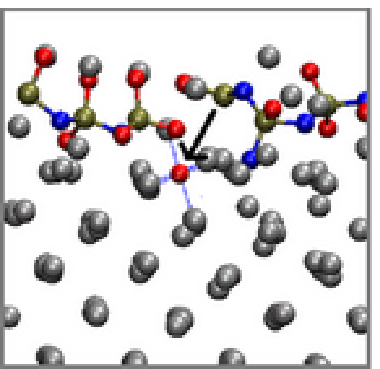} 
	           \epsfxsize=2.00in \epsfbox{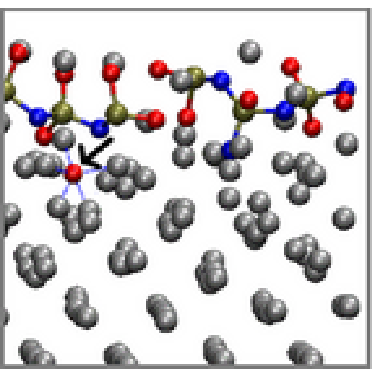} (f) }}
\caption[]
{\label{fig5} \noindent
(a) LiPON chain with one P-O-P linkage on Li metal; (b) broken P-O bond
along the backbone; (c) a different broken P-O bond along the backbone;
(d) breaking a P-N bond in (c); (e)-(f) breaking two different P-O bond in (c).
For color scheme, see Fig.~\ref{fig1}.
}
\end{figure}

Fig.~\ref{fig5}a depicts a P-O-P sequence in the P-N-P chain on Li surface.
It is derived by switching an N and an O atom in Fig.~\ref{fig3}a.  The total
energy of this unreacted chain is 1.56~eV higher than of Fig.~\ref{fig3}a,
which has the same stochiometry.  This shows that Holzwarth {\it et al.}'s
all N-P-N-backbone model is energetically far more favorble, and suggests
that P-O-P configurations generated under the fabrication conditions are
highly metastable.  Fig.~\ref{fig5}b represents a configuration where one P-O
bond, originally of length 1.65~\AA, is broken.  For this reaction,
$\Delta E$=-2.48~eV, and $\Delta E^*$ is only 0.90~eV.  The other P-O bond
in the backbone has an equilibrium bond length of 1.77~\AA.  Cleaving this
bond yields $\Delta E$=-3.14~eV, and $\Delta E^*$ is only 0.43~eV
(Fig.~\ref{fig5}c).  In both cases, Bader analysis\cite{bader} qualitatively
indicates that the 3-coordinated P atoms resulting from breaking the bond
are in +3 formal charge states.  These calculations accomplish our goal
of demonstrating the existence of at least one primary reaction that can
occur within a one-hour time frame at room temperature.

Fast LiPON reaction with Li metal is also observed in
Ref.~\cite{albe2017}, where the model LiPON backbone contains both
P-O-P bonds and N-(P)3 motifs, although the anode voltage is not reported
there.  We have not attempted inserting Li atoms into the {\it bulk}
LiPON region, unlike Ref.~\cite{albe2017}; this approach may be more
directly relevant to Li-vapor deposition experimental conditions reported
in XPS measurements\cite{jaegermann15,albe2017} rather than electrochemical
interfaces which are the focus of our calculations.

Next, we consider possible reactions subsequent to the formation of the 
structure in Fig.~\ref{fig5}c.  Fig.~\ref{fig5}d depicts the breaking of a
P-N bond where N is not part of the backbone.  The reaction is endothermic by
0.72~eV, and should not proceed.  Fig.~\ref{fig5}e depicts the cleavage of
an O-atom from 3-coordinated P$^{3+}$ atom.  This step is exothermic by
0.47~eV.  The predicted barrier is lower than those associated with
breaking other P-O bonds we have reported earlier, but remains a substantial
$\Delta E^*$=1.18~eV.  This magnitude for $\Delta E^*$ is consistent with a
reaction time scale that is still far beyond 1-hour at room temperature.  
Finally, Fig.~\ref{fig5}f depicts breaking a P-O bond on the -N-PO$_3$
terminus.  The reaction is exothermic by 0.28~eV, but the barrier
($\Delta E^*$=1.74~eV) is again high.  This last prediction dovetails with
our finding that that 4-coordinated N-PO$_3$ end groups in short LiPON
fragments are kinetically inert at room temperature (S.I.~Sec.~S3).

From these calculations, we conclude that LiPON can react with Li metal in a
one-hour time frame at room temperature, by cleaving P-O bonds within metasable
P-O-P sequences in the backbone.  After the initial bond-breaking, the
undercoordinated P$^{3+}$ atom is slightly more reactive; subsequent P$^{3+}$-O
bond breaking exhibits lower barriers than P$^{+5}$-O or P$^{+5}$-N linkages,
but the reaction rates associated with such reactions remain low compared to
battery operation timescales.  Surprisingly, despite the thermodynamic
instability of LiPON against lithium\cite{mo}, LiPON without P-O-P or N(P)$_3$
is kinetically robust on Li surfaces.  In contrast, C-O bonds at
Li$_2$CO$_3$/Li(s) interfaces exhibit far lower $\Delta E^*$, likely due to the
fact that the C atoms there are only 3-coordinated and have empty $p$ orbitals.

\section{Results: LiPON/Li$_x$CoO$_2$ (104) Interface}
\subsection{Explicit Interface}

\begin{figure}
\centerline{\hbox{ (a) \epsfxsize=1.80in \epsfbox{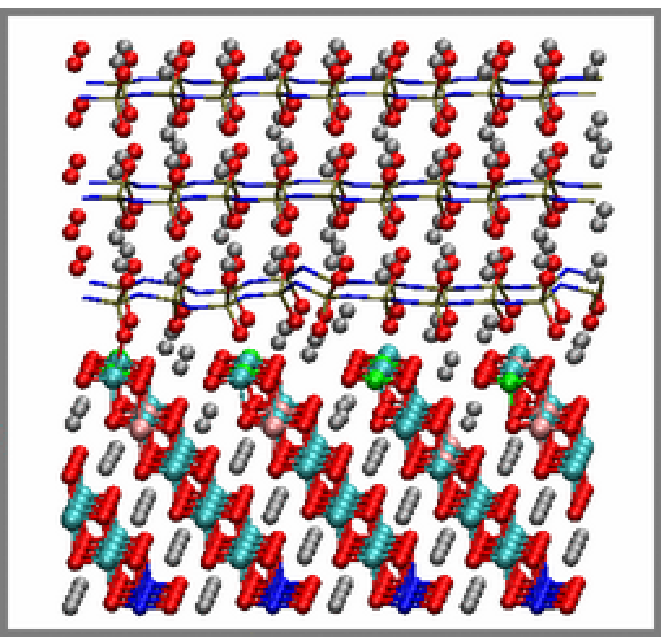} 
	           \epsfxsize=1.80in \epsfbox{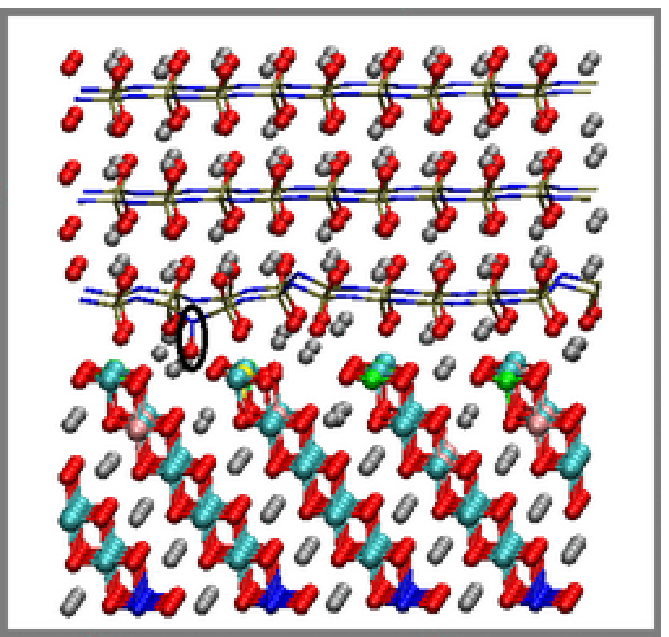} (b) }}
\centerline{\hbox{ (c) \epsfxsize=1.80in \epsfbox{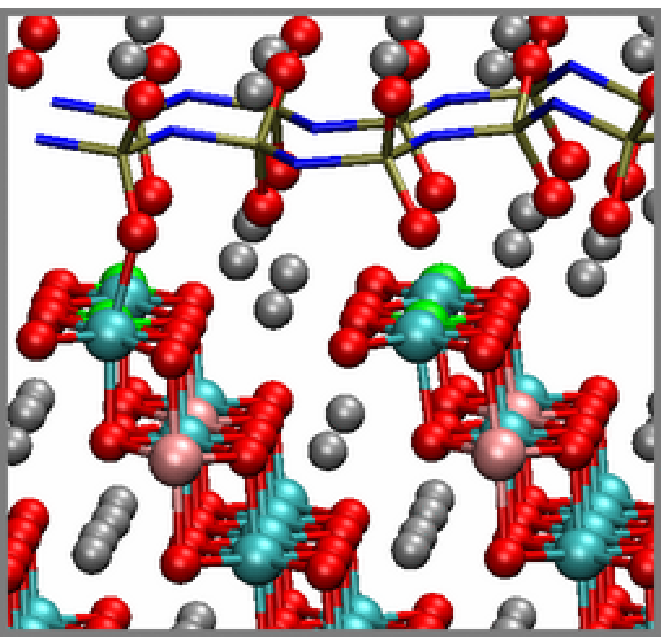} 
	           \epsfxsize=1.80in \epsfbox{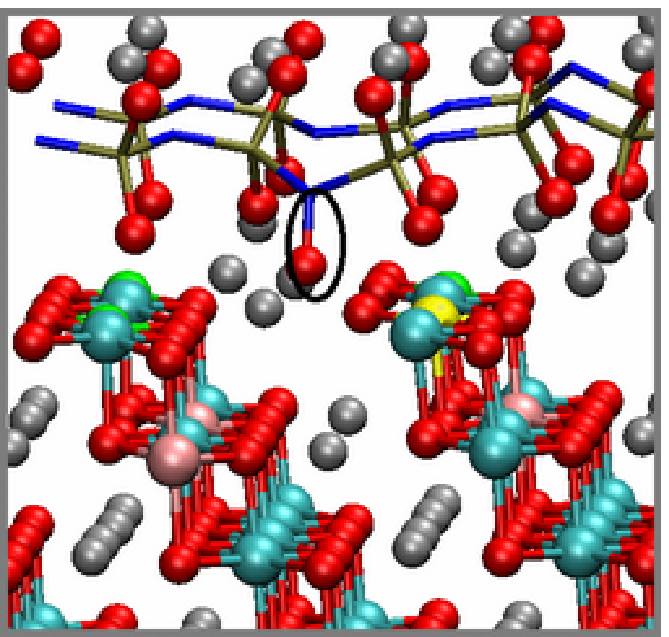} (d) }}
\centerline{\hbox{ (e) \epsfxsize=1.80in \epsfbox{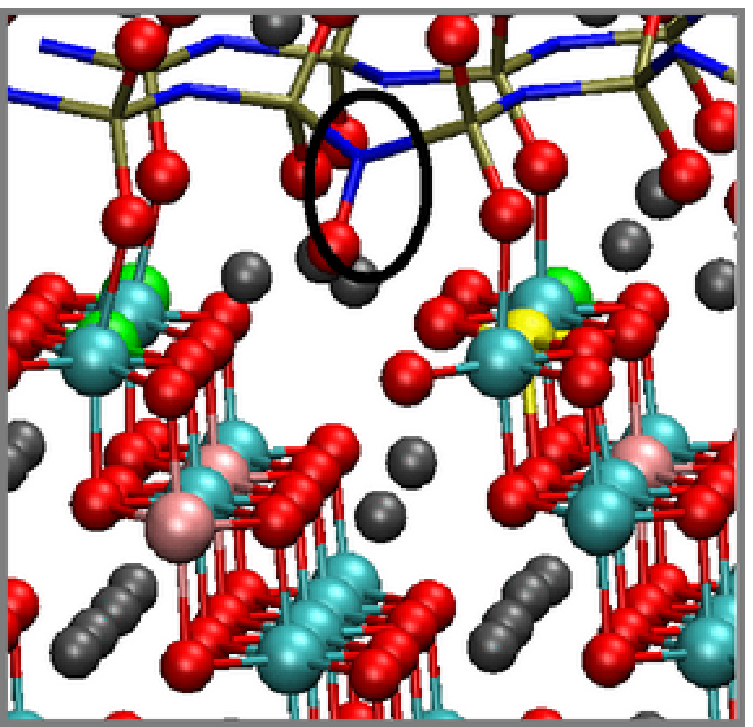} 
	           \epsfxsize=1.80in \epsfbox{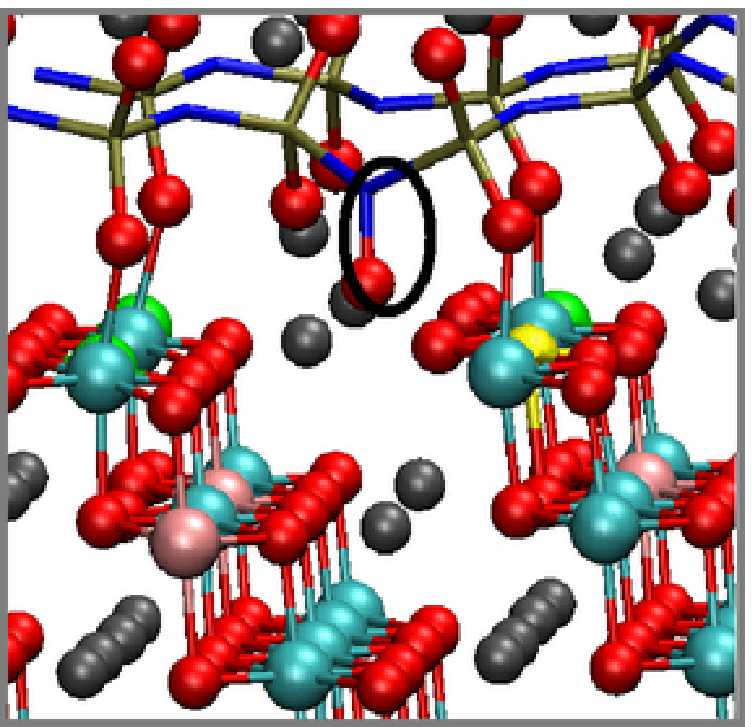} (f) }}
\centerline{\hbox{ (g) \epsfxsize=1.80in \epsfbox{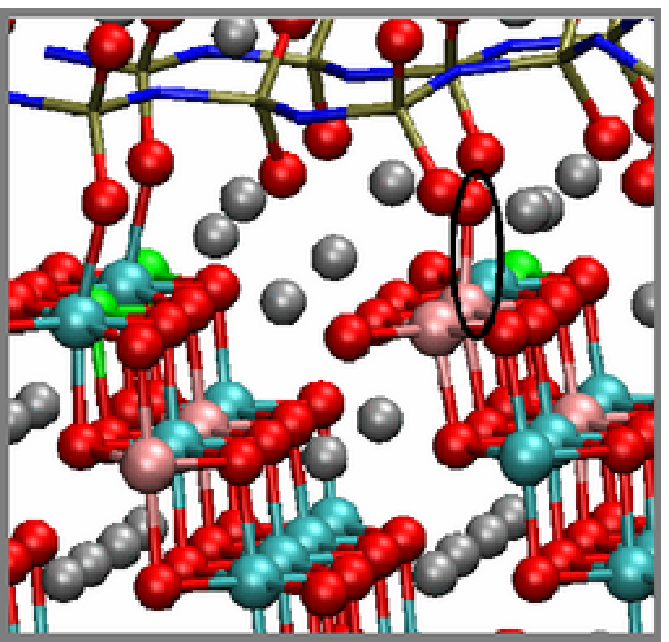} 
	           \epsfxsize=1.80in \epsfbox{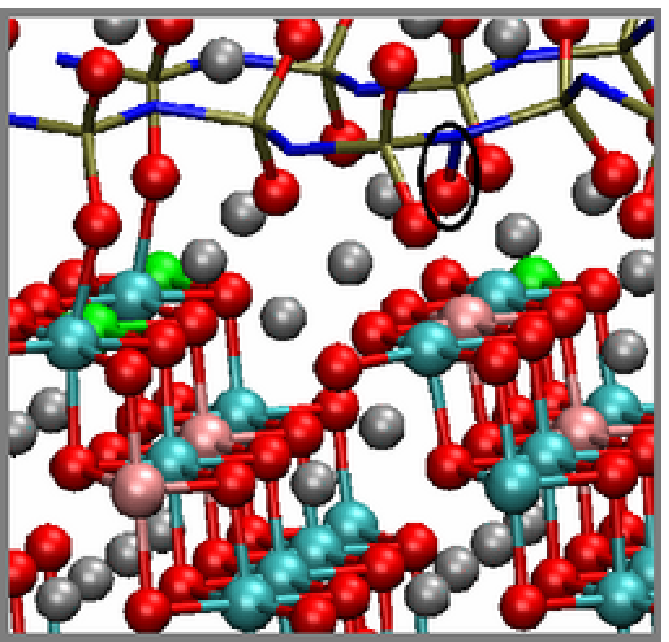} (h) }}
\caption[]
{\label{fig6} \noindent
(a)-(b) Original and reacted LiPON slab on Li$_x$CoO$_2$ (104) surface.
Circle indicates the formation of a N-O bond.  
(c)-(d) Same as (a)-(b), respectively, in expanded views.
(e)-(f) Similar to (d), but with distinct Li$^+$ removed from
the reaction zone. (g)-(h) Similar to (c)-(d), but with an extra Li$_2$O
unit added to the interface.  Low and high spin Co$^{4+}$ are depicted
as pink and green spheres.  Low, intermediate, and high spin Co$^{3+}$
are depicted cyan, blue, and yellow spheres.  The rest of the color scheme is
as in Fig.~\ref{fig1}, but P and N atoms are now depicted as sticks, not
spheres.
}
\end{figure}

This section focuses on the interface between model LiPON and
Li$_x$CoO$_2$ (104) (Fig.~\ref{fig6}a).  The predictions herein are more
qualitative, partly because of uncertainties in the voltages associated with
the simulation cells.  

Small $x$ in Li$_x$CoO$_2$ is consistent with high equilibrium voltage,
which should increase degradation\cite{meng1}.  In this work, $x$
is set to a fairly large value, 0.83, to facilitate convergence of DFT
calculations.  (See the Technical section and the S.I., Sec.~1,
for details.) The cobalt spin states are also depicted in Fig.~\ref{fig6}.
Cobalt exhibits low-spin Co$^{3+}$ and low-spin Co$^{4+}$ states in
the interior of the cathode slab.  Half the cobalt at the interface are
bonded to LiPON O~atoms; they are 6-coordinated low-spin Co$^{3+}$.
In contrast, 5-coordinated cobalt ions on the surfaces are in intermediate-spin
Co$^{3+}$\cite{mengsur,sumita} and high-spin Co$^{4+}$ states.  
There are a total of 63~net up-spin in these simulation cells.  Switching
to 65 net unpaired electrons changes $\Delta E^*$ by only 0.025~eV.

Fig.~\ref{fig6}a-d depict the configurations before and after moving an O atom
from the Li$_x$CoO$_2$ surface to a LiPON N-atom at the interface.  The N-O
distances in these configurations are 2.36~\AA\, and 1.35~\AA, respectively.  
The distance between the reacted N~atom and one of the two P-atoms to which it
is bonded increases from about 1.65~\AA\, to 1.75~\AA.  The 5-coordinated
Co$^{4+}$ on the surface originally bonded to the transferring O$^{2-}$ in 
Fig.~\ref{fig6}c, colored in green, has turned into a 4-coordinated, 
Co$^{3+}$ (yellow) in Fig.~\ref{fig6}d.  It now has four unpaired
electrons, reminiscent of 4-coordinated Co$^{3+}$ on the stoichiometric
(110) surface\cite{mengsur}.  Another low-spin, 6-coordinated Co$^{4+}$ in
the second oxide layer has become a low-spin Co$^{3+}$.  These changes
are consistent with the loss of two $e^-$
from the transferring O$^{2-}$ to LCO.  In other words, LiPON has been
oxidized.  As further confirmation, maximally localized Wannier orbital
analysis shows that, in the P-(N-O)-P group created, a charge-neutral O atom
is transferred, forming a dative covalent bond with the N-atom.  In contrast,
oxygen atoms on the unreacted cathode surface have -2~formal charges.  

$\Delta E$ and $\Delta E^*$ for this reaction are +0.29~eV and +1.09~eV,
respectively.  This indicates that the O-transfer reaction is thermodynamically
unfavorable and kinetically slow.  

Increasing the cathode voltage should favor oxidation of LiPON.  Rigorously
speaking, raising the voltage (${\cal V}_e$) requires lowering the cathode Fermi
level.  This should be accompanied with removal of Li$^+$ and $e^-$ pairs in a
consistent, grand canonical ensemble manner; Li atoms with vacancy formation
energies below $|e| {\cal V}_e$ should be removed from the simulation
cell.  As Li$_x$CoO$_2$ is a polaronic conductor with no band gap in parts of
the phase diagram, the simplest way to model voltage dependence is to add
a metallic ``current collector''\cite{solid}.  However, in our case, net
spin can accumulate in the metal slab, which would hinder the control of
the total spin in LCO.  Determining which Li atom(s) to
remove in the interfacial region is also a difficult task, given the disordered
configuration there; removal of many distinct interfacial Li atoms have to be
attempted.  Here we make a local approximation.  We remove one Li$^+$ plus
an $e^-$ from the system, and re-compute $\Delta E$ and $\Delta E^*$.  
Rigorous voltage determinations\cite{solid} are deferred to future work.

Fig.~\ref{fig6}e-f are obtained by removing one Li from Fig.~\ref{fig6}d.  
They represent two Li deletion choices, and exhibit three Li in the reaction
zone instead of four Li in Fig.~\ref{fig6}d.  The pre-reaction configurations
are similar to Fig.~\ref{fig6}a and are not shown; compared with
Fig.~\ref{fig6}c, they both entail an energy cost of 4.30~eV after accounting 
or the chemical potential of the Li removed.  While not rigorous, this suggests
that the ``voltage'' associated with Li loss is about 4.30~V vs.~Li$^+$/Li(s)
in both cases before O-transfer, if the Li-content is indeed at equilibrium
with the instantaneous ${\cal V}_e$.  $\Delta E$ associated with O~transfers
from the LCO surface to LiPON are predicted to be +0.11 and +0.10~eV, while
$\Delta E^*$ are also almost indistinguishable 0.98~eV and 0.95~eV.

In general, $\Delta E$$<$0 is required for reactions to go forward.  But further
lowering the $x$ value further below our current $x$=0.83 is expected to be
consistent with more negative, ultimately favorable $\Delta E$.  Section~S4
of the S.I.~further suggests that the DFT+U method used in this
work overestimates $\Delta E$; a more generally accurate functional like PBE0
should reduce $\Delta E$ and render the reaction exothermic.  Finally,
in Sec.~S5 of the S.I., we show that a similar O-transfer reaction between
this crystalline LiPON model and the Li$_x$CoO$_2$ (110) surface is 
exothermic by 0.36~eV even when using the DFT+U method.  Regarding the  
barrier, $\Delta E^*$ are 0.95 and 0.98~eV in the two panels, consistent with
reaction times of roughly one hour at room temperature.  Therefore we assert
that this set of calculations show that interfacial reactions between the
cathode and the electrolyte are viable at room temperature.  These predictions
are consistent with apparent oxygen loss from Li$_x$CoO$_2$ -- especially
during charging at high temperature\cite{meng1,meng2}. The change in
spin polarization predicted at the LiPON/Li$_x$CoO$_2$ interface may be
measurable.  Note that our calculations pertain to high equilibrium voltages,
not as-grown conditions.  

Section~S4 of the S.I. demonstrates that moving an O~atom from the bulk
(as opposed to the surface) of LCO to LiPON is energetically more unfavorable.
This trend is generally observed on cathode oxide surfaces\cite{persson}.
Hence continuous loss of O~atoms from LCO must be mediated by other mechanisms. 
One possibility is the migration of undercoordinated surface Co from their
surface sites, which creates more undercoordinated oxygen at LCO surfaces.

So far we have focused on flat Li$_x$CoO$_2$ (104) surfaces.  Other models, in
which (104) surfaces are covered with CoOH groups due to reaction with H$_2$O
in the atmosphere, have
been proposed\cite{musgrave}.  Fig.~\ref{fig6}g-h explore this possibility
by adding a Li$_2$O formula unit at the interface, with the added O$^{2-}$
attached to a formerly 5-coordinated Co ion.  Using computational procedures
similar to those used above, we find that $\Delta E$=-1.61~eV and
$\Delta E^*$=0.10~eV for transferring the newly added O~atom from the Co
ion to a LiPON nitrogen atom nearby.  The reaction proceeds much more readily
since there is no need to create an oxygen vacancy on the cathode surface.
This calculation strongly suggests that no Co-O bond with O~atom sticking
out of the surface survives contact with LiPON.

\begin{figure}
\centerline{\hbox{ (a) \epsfxsize=1.50in \epsfbox{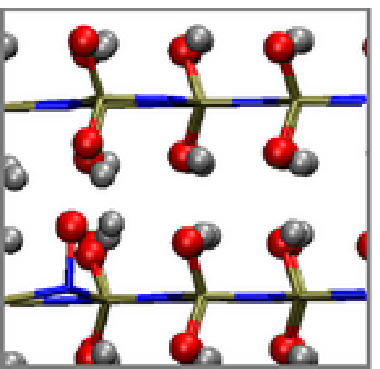} 
	           \epsfxsize=1.50in \epsfbox{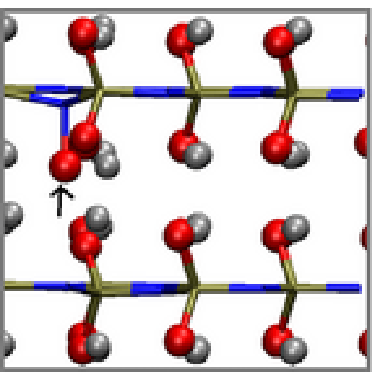} (b) }}
\centerline{\hbox{ (c) \epsfxsize=1.50in \epsfbox{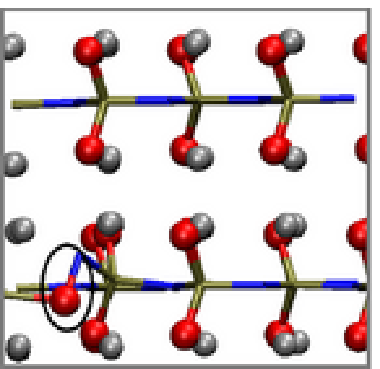} 
	           \epsfxsize=1.50in \epsfbox{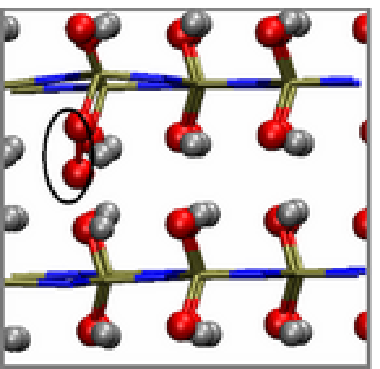} (d) }}
\caption[]
{\label{fig7} \noindent
Li$_2$PO$_2$N bulk crystal simulation cell with eight formula units and one 
extra O~atom bonded to a N-atom in the backbone.  (a) and (b) are two
topologically equivalent configurations with N-O bonds and have equal
energies.  The text describes the transfer of the extra O~atom between them.
(c) O~atom inserted between P and N.  (d) O~atom added to a LiPON O~atom.
}
\end{figure}

\subsection{Excess Oxygen Diffusion Inside LiPON}

Next we consider possible subsequent oxygen migration steps.  Here we turn
to LiPON bulk ``crystal'' models not in contact with Li$_x$CoO$_2$ to
represent interior LiPON regions away from the interface.  This is done to
reduce the computational cost.  Fig.~\ref{fig7}a depicts a LiPON crystal
supercell with eight formula units and one O~atom added to one of the LiPON
N~atoms, like in Fig.~\ref{fig6}d.  Fig.~\ref{fig7}b depicts another optimized
configuration where the added O~atom has been manually moved to an equivalent
position on a neighboring LiPON chain.  The two configurations have identical
energies.  The diffusion barrier between them is a low 0.68~eV.  This suggests
that O~atoms abstracted from Li$_x$CoO$_2$ readily diffuse inside LiPON
once it is away from the interface, possibly aiding further creation of
oxygen vacancies on the cathode surface.

We also explore other locations where an additional O~atom can insert
into LiPON.  Fig.~\ref{fig7}c depicts a P-O-N motif which is 0.17~eV more
favorable than the P-N-O linkage in Fig.~\ref{fig7}a-b.  This configuration
can be a reaction product or intermediate subsequent to N-O bond formation in
Fig.~\ref{fig7}a-b.  In contrast, O-O bond formation (Fig.~\ref{fig7}d) is
less favorable than Fig.~\ref{fig7}a-b by 0.64 eV.  N-O bond formation appears
a crucial step in LiPON oxidation by LCO.  We predict that that N~atoms are
the reactive sites at LiPON/cathode interfaces, and propose that a lower
N-content at the interface may provide better kinetic LiPON stability against
Li$_x$CoO$_2$.

While various defects have been considered in LiPON
simulations\cite{holzwarth2008}, to our knowledge there has been little effort
to model or measure the migration of excess oxygen bonded to N-atoms in the
LiPON backbone.  It is possible that the amorphous arrangements of LiPON
chains in experimental samples may impede interchain oxygen transport.

\section{Results: Conductivity and XPS}

\begin{figure}
\centerline{\hbox{ \epsfxsize=4.00in \epsfbox{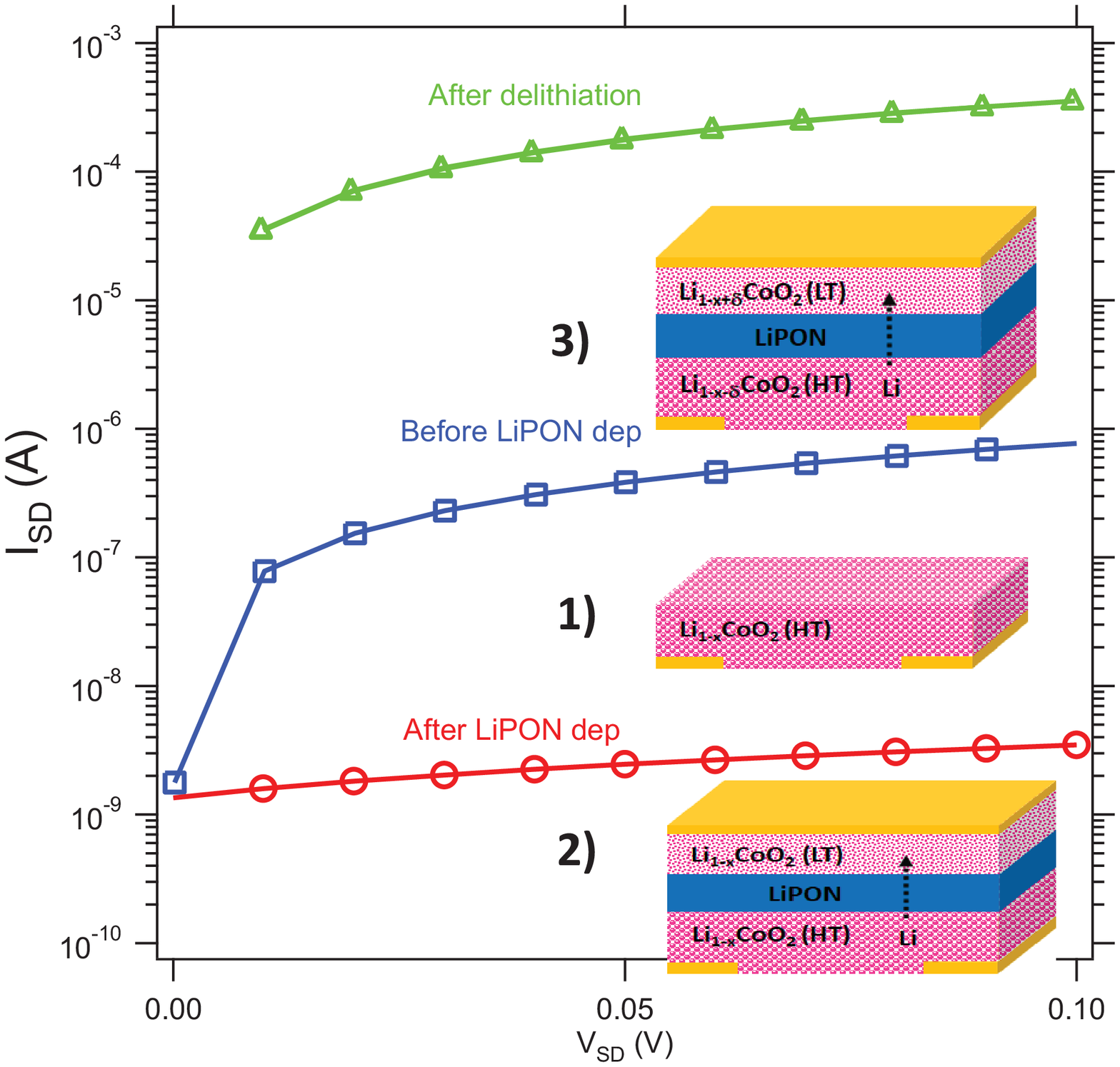} }}
{\label{fig8} \noindent
Fig.~8.  Current ($I_{\rm sd}$) as a function of applied voltage ($V_{\rm sd}$)
in solid-state devices.  Blue: bare LiCoO$_2$;  red: LiCoO$_2$ after
LiPON deposition; green: same device as depicted in red but with an
applied bias that delithiates LiCoO$_2$.  (1)-(3), in that order, depict
currents associated with the sequence of events described in the text.
}
\end{figure}

This section reports some conductivity and XPS measurements relevant to 
the LCO/LiPON interface calculations above.  Experimental LiPON is 
amorphous, unlike the crystalline models.  However, the local structures
responsible for chemical reactions are not expected to depend on LiPON
long-range order.

The electronic conductivity of LiCoO$_2$ varies by as much as six orders of
magnitude depending on Li concentration\cite{delmas,milewska2014,fuller2017},
and can be used as a sensitive probe of lithitation\cite{fuller2017}.
To experimentally investigate the transfer of Li between LiCoO$_2$ and a
LiPON electrolyte, electrochemical transistors were fabricated consisting
of a thin film stack of sputtered LiCoO$_2$/LiPON/LiCoO$_2$ layers, with the
bottom LiCoO$_2$ acting as a transistor channel and the top LiCoO$_2$ acting
as an electrochemical gate.  Sputtered LiPON is known to yield amorphous
LiPON\cite{bates1993}.  The fully fabricated transistor cell consists
of 100~nm of high-temperature (HT) LiCoO$_2$, 400 nm of LiPON, 100~nm of 
low-temperature (LT) LiCoO$_2$. The transistor channel was fabricated using
photo lithographically defined Pt electrodes (60~nm) with channel dimensions of
4~$\mu$m length and 1700~$\mu$m width. Further details of device fabrication
process were described previously\cite{fuller2017}. 

The electronic conductivity of the bottom LiCoO$_2$ layer was measured
after deposition of each subsequent layer in order to qualitatively
understand the Li transfer during sputter fabrication processes.

Figure 8 depicts the current vs. voltage characteristics of the
transistor source drain terminals with an illustration of the
various layers at the time of measurement. The current is measured
parallel to the LiCoO$_2$ surface and/or the LiCoO$_2$/LiPON interface.
Initially (1), there is no external bias normal to the surface. At the
same applied voltage parallel to the interface, the current is much higher
before LiPON deposition (1) than after (2), indicating that the resistivity
has gone up significantly. This is consistent with the transfer of Li
(i.e., Li$^+$ and $e^-$.) from LiPON to LiCoO$_2$. The latter may be slightly
Li-defective when first synthesized. Lithiation of Li$_x$CoO$_2$ at $x$$\sim$1
is known to increase its resistivity.  This point is confirmed by
applying an external bias perpendicular to the LiCoO$_2$/LiPON interface,
which removes Li from LiCoO$_2$ to the electrochemical gate on the
other side of the LiPON film (3). The magnitude of the current (green line)
goes back up, to a value above that obtained before LiPON deposition.

The results depicted in Fig.~8 are consistent with removal of Li$^+$ and $e^-$
from LiPON.  By themselves they do not yield evidence of oxygen migration from
LiCoO$_2$ to LiPON.  But LiPON oxidation is most readily accomplished by
oxidizating either O- or N~anions.
Li$^+$ and P$^{5+}$ atoms cannot be further oxidized.  Adding O to LiPON
would be consistent with our predictions in the last section.  Note that the
excess Li in LiCoO$_2$ could also be the cause of the disorder observed in
Refs.~\cite{meng,meng1,meng2}.  One possible secondary reaction
after the oxygen transfer is a charge-neutral NO molecule release from P-(NO)-P
created in the first step, with the LiPON surface losing an $e^-$ and a Li$^+$
to Li$_x$CoO$_2$ in the processs.  This suggestion from our experimental
collaboration will be considered in future computational work.

\begin{figure}
\centerline{\hbox{ \epsfxsize=5.00in \epsfbox{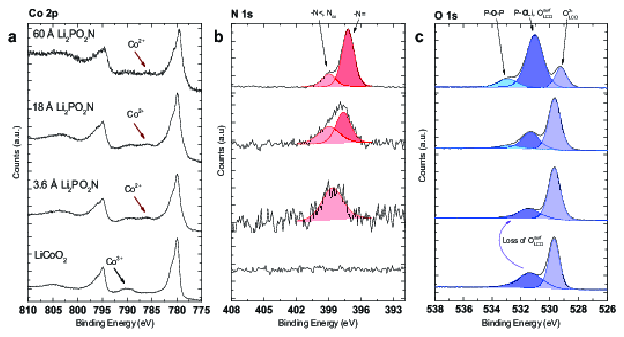} }}
{\label{fig9} \noindent
Fig. 9.  XPS spectra for various thicknesses of LiPON films deposited on
LiCoO$_2$.  (a) Co region; (b) N; (c) O.
}
\end{figure}

We have also performed XPS measurements on LiPON deposited on LiCoO$_2$ using
the atomic layer deposition (ALD) method, prior to cycling.  Our previous
studies have shown that this method yields amorphous LiPON in the tested
temperature range, as indicated by the lack of identifiable peaks in
X-ray diffraction\cite{lipon3}.  The results are
depicted in Fig.~9.  It shows the Co $2p$, O $1s$, and N $1s$ spectra of
0, 6, 30, and 100 cycles of ALD Li$_2$PO$_2$N grown at T=250$^o$C, resulting
in nominally 0, 3.6, 18, and 60~\AA~of coverage over LCO, respectively.
The deposition process immediately produces a satellite feature in the
Co $2p$ spectrum consistent with reduction of surface states to Co$^{2+}$.
The O~1s spectrum shows a loss in what is often considered ``surface'' or
``undercoordinated'' O feature in Li$_x$CoO$_2$, though this is not a strong
definition.  It could be consistent with loss of these oxygen atoms to the ALD
precursors/Li$_2$PO$_2$N layer, as the N $1s$ spectrum clearly shows that the
N atoms closest to the interface are all in a more highly oxidized state than
in the ``bulk'' Li$_2$PO$_2$N. This is similar to what other groups have seen
with other methods of growing LiPON\cite{jaegermann10,hausbrand}.  The
N-O bond formation predicted in Fig.~\ref{fig6} is consistent with oxidized
nitrogen.  There is a clear signature of P-O-P sequence in the O~spectra.
This feature is reflected in our anode model depicted in Fig.~\ref{fig5}. 

Note that some of these features, especially Co$^{2+}$, are at least
partially produced by vacuum annealing of LCO at 250$^o$C.  Co$^{2+}$ is
not seen in our calculations because the partial delithiation in our DFT
models should yield Co$^{3+}$ and Co$^{4+}$.  We also stress that the samples
and devices analyzed in Fig.~9 are distinct from those described in Fig.~8.

\section{Discussions: Comparison with Experimental Literature}

Our interfacial simulations focus on initial reaction barriers in 
crystalline models.  They are not meant to predict final chemical
speciations.  Hence comparisons with structural measurements require
some extrapolation -- especially since experimental LiPON is amorphous.
The previous section has revealed qualitative agreement between our
predictions and measurements.  This section focuses on comparison with
published experimental work.

As stated in the Introduction, all-solid-state batteries with Li metal anode,
LiCoO$_2$ cathode, and LiPON electrolytes have been shown to cycle 
well if the LiPON film is sufficiently thick, especially at room
temperature\cite{review4,lipon3,meng1,meng2}.  At the same time,
some degradation products and/or disordered regions are reported at both
cathode\cite{meng1,meng2,jaegermann10,hausbrand} and
anode\cite{jaegermann15,albe2017} interfaces.  While the specific LiPON
structure or composition used in experiments may not coincide with our models
or even with each other, our predictions can help interpret these results.

At the anode interface, we predict that only P-O-P sequences can readily
break at room conditions.  This mechanism is also proposed in
Ref.~\cite{jaegermann15}; see in particular Fig.~4 there.  Most other
bond-breaking barriers, e.g., those involving P-N cleavage, are predicted to
exceed $\Delta E^*$=1.5~eV.  This suggests that temperatures of $>$450~K, which
approach the Li melting point, are needed to make these P-N-P/Li(s) reactions
fast enough to occur at room temperature.  This finding is consistent with
the survival of LiPON XPS signals seen in vapor deposition of Li on
LiPON\cite{jaegermann15}.  The ratio between triply- and doubly-coordinated
N atom in amorphous LiPON significanty decreases upon deposition
Li,\cite{jaegermann15} consistent with the hypothesis that triply-coordinated
N, which does not exist in our model, is far more reactive than
doubly-coordinated N.  To the best of our knowledge, the Li/LiPON
interface has not been reported as a major source of degradation in
all-solid-state batteries.  Our predictions are also consistent with the
kinetic stability observed in Pt/LiPON/Li devices, which are stable for months
with little change in conductivity\cite{lipon3}, although some SEI products
from ALD-deposited LiPON likely also help passivation.

At the cathode surface, a 10~\AA\, thick NO$_2^-$ and/or NO$_3^-$ layer has
been estimated from XPS measurements\cite{jaegermann10,hausbrand}.  The
formation of N-O bonds is consistent with the first step reaction mechanism
predicted in our calculations (Fig.~\ref{fig6}).

More recently, disordered LiCoO$_2$ regions in contact
with LiPON have been reported in STEM studies\cite{meng1,meng2}.
The disordered layer can be hundreds of nanometer thick at room temperature
prior to cycling.  Its thickness increases with charge/discharge and
especially with temperature.  However, the all-solid battery retains its
capacity after 250 cycles\cite{meng1}.  The LiPON region does not exhibit
significant changes, possibly because LiPON is already amorphous.  Co$_3$O$_4$
and Li$_2$O$_2$ are identified in the disordered Li$_x$CoO$_2$ region by STEM
and electron energy loss spectroscopy\cite{meng1,meng2}.  These measurements
suggest the presence of Co$^{2+}$ at the interface and imply loss
of oxygen from Li$_x$CoO$_2$.  Oxygen transfer from LiCoO$_2$ to LiPON are
proposed to occur already at the as-grown, uncycled LiCoO$_2$/LiPON interface
in these experimental works.  Peroxides species are also reported at these
interfaces\cite{meng}.  Recall that our XPS measurements (Fig.~9)
also indictate the presence of Co$^{2+}$.

Our calculations focus on charged Li$_x$CoO$_2$, $x=0.83$, which only contains
Co$^{3+}$ and Co$^{4+}$ ions.  No Co$^{2+}$ is expected at this $x$
value.  We predict loss of oxygen from Li$_x$CoO$_2$.  Upon transfer of an
O-atom to LiPON, two Co$^{4+}$ turn into Co$^{3+}$.  The oxygen
vacancies formed may be consistent with a disordered Li$_x$CoO$_2$
region\cite{meng}.  Our calculations focus on the initial stages of reaction,
and provide no information about the thickness of the reacted cathode layer.
We have not observed peroxide formation.  Peroxide-like structures have been
predicted in oxygen-depleted Li$_2$MnO$_3$\cite{henkelman}.  In our simulation
cells, however, N-O bond formation is found to be more favorable than O-O.

\section{Conclusions}

In conclusion, we have applied electronic structure DFT calculations to study
interfacial degradation reactions between model crystalline LiPON and the
surfaces of two electrodes: Li metal and Li$_x$CoO$_2$.  Some experimental
corroboration is also provided; this assumes that interfacial reactions on
amorphous and crystalline LiPON models are similar.  The predictions are
relevant to the
interfacial film (``solid electrolyte interphase'' or ``SEI'') products
formed during cycling of all-solid-state batteries using LiPON solid
electrolytes.  LiPON proves to be an interesting case study.  Single phase
thermodynamics calculations predict instabilities at both interfaces, which 
do not distinguish the extent of reactions formed in the two cases.  In this
work, we instead focus on models with explicit interfaces and
calculate the reaction activation energies.  Multiple reaction sites and
bond-breaking events are considered.  The predictions suggest that most
initial reactions on the anode surface are slow for LiPON with P-N-P backbone
and ALD-like stoichiometry,\cite{lipon3} while cathode
interfacial reactions can occur within battery cycling timescales.

On lithium (001) surfaces, ordered LiPON chains with P-N-P backbones, P-O
side groups, and 2-coordinated N atoms are found to exhibit P-N and P-O
cleavage barriers in excess of 1.4~eV, which correspond to reactant half
lives far in excess of battery operation timescales.  In contrast, P-O-P
sequences, much less energetically favorable but known to exist in LiPON, are
found to exhibit faster bond-breaking reactions.  However, subsequent reactions
again exhibit barriers exceeding 1~eV and are slow.  The electrode potential
does not strongly affect the exothermicity or the reaction barrier.  From 
these calculations, some SEI products are expected at this interface, but
extensive degradation is not expected at room temperature.  
This is consistent with experimental data showing that some 2-coordinated
N persists after lithium vapor deposition.\cite{jaegermann15}  Our model
thus helps pinpoint less reactive LiPON motifs. This is consistent with
experimental data. Our predicted reaction rates are much lower than those
in models with both N(P)$_3$ and P-O-P groups.\cite{albe2017}

At LiPON/Li$_x$CoO$_2$ (104) interfaces, cobalt ions can exhibit five
different spin/charge states, which makes reproducible calculations of
Kohn-Sham wavefunctions difficult.  We believe this is a general phenomenon
associated with Li$_x$CoO$_2$ surfaces, and propose that extra care should be
taken in future modeling of interfacial spin states associated with this
material.  By working close to $x$=1, and propagating wavefunctions
quasi-continuously from product to reactant, estimates for reaction
barriers are obtained.  We find that even the surfaces of
chemically ordered, crystalline LiPON slabs are oxidized by Li$_x$CoO$_2$
within battery cycling (one-hour) timescale at room temperature.  The LiPON
N-atom abstracts a O~atom from the oxide surface in the process.  O-atoms
added to crystalline LiPON interior are mobile, potentially creating pathways
for further degradation of Li$_x$CoO$_2$ as battery cycling continues.
We propose that interfaces with less nitrogen content may yield less
degradation on cathodes.

This work emphasizes kinetics, not thermodynamics, at solid-solid interfaces.
Under processing ($\sim$200~$^o$C) and cycling (room temperature) conditions,
electrode-electrolyte interfaces may not be at thermodynamic equilibrium, and
metastable starting materials and/or intermediate products may persist.
In addition to shedding light on reaction kinetics, our calculations
elucidate the low-barrier initial bond-breaking steps involved in 
degradation reactions.  This will facilitate future design of solid
state materials and interfaces more resilient to degradation.

\section{Technical Section} \label{method}

DFT calculations are conducted using the Vienna Atomic Simulation Package
(VASP) version 5.3\cite{vasp,vasp1,vasp2} and the PBE functional\cite{pbe}.
Modeling Li$_x$CoO$_2$ with $x\sim$$1$ requires spin-polarized DFT+U augmented
treatment\cite{dftu} of Co $3d$ orbitals.  The $U$ and $J$ values depend on the
orbital projection scheme and DFT+U implementation details; here $U-J=$3.30~eV
is adopted in accordance with the literature\cite{cedersur}. In Sec.~S4 of
the S.I., the more computationally costly PBE0 functional\cite{pbe0}, which
is generally more accurate, is used for spot-checks.  

We adopt one of the crystalline LiPON (Li$_2$PO$_2$N ``s2'') crystal structures
created by the Holzwarth group\cite{holzwarth_bulk}.  This model consists of
parallel zig-zag LiPON chains, and is chosen because the lattice dimensions,
re-optimized using the PBE functional, best match the Li (001) supercell size.

Our interfacial model systems are charge-neutral asymmetric slabs.  The details
of representative simulation cells are listed in Table~\ref{table1}. The
standard dipole correction is applied to negate image interactions in the
periodically replicated, charge-neutral anode-side simulation
cells\cite{dipole_corr}.  This correction is found to be O(1)~meV on the
cathode side, and is omitted therein.

Reaction barriers are computed using the climbing-image nudged
elastic band (cNEB) method\cite{neb}.  The barriers associated with
LiPON P-N bond cleavage on lithium surfaces are non-trivial to compute because
cNEB can mistake (P-O)-Li$^+$ dissociation with true bond-breaking events.
When the cNEB approach yields a configuration close to the barrier top, we
typically switch to quasi-Newton algorithm optimization of that single
configuration until the maximal force on each atom is less than 0.07~eV/\AA.
When Li$_x$CoO$_2$ slabs are present in the simulation cell, we have
propagated wavefunctions quasi-continuously from product to reactant.
More details on computational and experimental methods
are found in Sec.~S1 of the S.I.

\begin{table}\centering
\begin{tabular}{c|r|r|l|r} \hline
system & dimensions & stoichiometry & $k$-sampling & Figure \\ \hline
Li(s)/LiPONc &  14.28$\times$16.41$\times$28.00 & 
	Li$_{172}$P$_{6}$O$_{12}$N$_{6}$ &
          2$\times$2$\times$1  & Fig.~3a \\
Li(s)/LiPONf &  14.28$\times$23.06$\times$28.00 & 
	Li$_{236}$P$_{6}$O$_{12}$N$_{6}$ &
          2$\times$1$\times$1  & Fig.~S1a \\
Li(s)/LiPONs &  14.28$\times$16.41$\times$40.00 & 
	Li$_{274}$P$_{54}$O$_{108}$N$_{54}$ &
          2$\times$2$\times$1  & Fig.~1a \\
%LCO/LiPONs half &  5.71$\times$23.62$\times$36.00 & 
%	Li$_{100}$Co$_{48}$O$_{156}$P$_{30}$N$_{30}$ &
%          2$\times$1$\times$1  & Fig.~1a \\
LCO/LiPONs &  11.41$\times$23.62$\times$36.00 & 
	Li$_{195}$Co$_{96}$O$_{312}$P$_{60}$N$_{60}$ &
          2$\times$1$\times$1  & Fig.~6a \\
LiPON &  10.94$\times$9.26$\times$9.52 & 
	Li$_{32}$P$_{16}$O$_{32}$N$_{16}$ &
          2$\times$2$\times$2  & Fig.~7a \\
Li$_x$CoO$_{2-\delta}$ &  8.53$\times$8.53$\times$14.18 (hex)  & 
	Li$_{25}$P$_{27}$O$_{53}$ &
          2$\times$2$\times$1  & NA \\
\hline
\end{tabular}
\caption[]
{\label{table1} \noindent
Computational details of representative simulation cells.  LiPONs, LiPONc,
and LiPONf refer to a LiPON slab, a single chain, and a fragment respectively.
Typically the configurations are first optimized using $\Gamma$-point sampling
and then re-optimized using the listed $k$-point grid.
}
\end{table}

Lithium is a metallic conductor and its Fermi level ($E_{\rm F}$) is well
defined.  Work functions are computed as differences between $E_{\rm F}$ and
vacuum levels.  The work function minus 1.37~V is the instantaneous electronic
voltage (${\cal V}_e$) relative to Li$^+$/Li(s).  We distinguish ${\cal V}_e$
from the equilibrium or ionic voltage, which is function of the lithium chemical
potential referenced to lithium metal cohesive energy\cite{solid}.  There
is no reason to expect that DFT interfacial models are automatically at
electrochemical equilibrium, in the sense that the two definitions are
equal.  As discussed in the text, such models are more likely to be at
overpotential conditions unless care is taken.

The charge states of Co ions are determined by
cross-referencing maximally localized Wannier function analysis\cite{wannier}
and approximate local spin polarzations $s_z$ predicted by the VASP code.
$|s_z|$$\approx$0.0, 1.0, 2.0, 2.2, and 2.8 are assigned to low-spin 
Co$^{3+}$, low-spin Co$^{4+}$, intermediate-spin Co$^{3+}$,
high spin Co$^{4+}$, and high-spin Co$^{3+}$, respectively.  Note
that the VASP code requires that Wannier calculations be conducted using
$\Gamma$-point Brillouin zone sampling.

\section*{Acknowledgement}

We thank Yue Qi for useful discussions.  
This work was performed, in part, at the Center for Integrated
Nanotechnologies, an Office of Science User Facility operated for the U.S.
Department of Energy (DOE) Office of Science.  It was supported
by Nanostructures for Electrical Energy Storage (NEES), an Energy Frontier
Research Center funded by the U.S.~Department of Energy, Office of Science,
Office of Basic Energy Sciences under Award Number DESC0001160.  
Sandia National Laboratories is a multimission laboratory managed and
operated by National Technology and Engineering Solutions of Sandia, LLC,
a wholly owned subsidiary of Honeywell International, Inc., for the
U.S.~Department of Energy’s National Nuclear Security Administration under
contract DE-NA0003525.


\begin{thebibliography}{99}
\bibitem{note1}
Operationally, assembling interfaces often involves atomic layer deposition,
plasma laser deposition, and other technique to deposit LiPON or other
solid electrolytes on to electrode surfaces.   These techniques do not
correspond to pressing the flat surfaces of two pre-formed materials together.
Intermixing of electrolye and electrode components at the interface may occur
naturally during growth; therefore the SEI interlayers may result from
gas phase reactions in addition to degradation of fully-formed LiPON.  What
is needed to address such growth conditions is a comprehensive science of
rocessing conditions.
\end{thebibliography}
\end{document}